\documentclass[12pt]{book}

\usepackage[dvips]{graphicx,color}
\usepackage{makeidx,tsukuba}

\makeauthorindex
\makeindex

\begin{document}

\BookTitle{\itshape The 28th International Cosmic Ray Conference}
\CopyRight{\copyright 2003 by Universal Academy Press, Inc.}
\pagenumbering{arabic}

\chapter{
Processing of the Signals from the Surface Detectors of the Pierre Auger 
Observatory}

\author{%
%
%
T.Suomij\"{a}rvi$^1$, Pierre Auger Collaboration$^2$ \\
{\it (1) Institut de Physique Nucleaire, IN2P3-CNRS, Universit\'e Paris-Sud, 91406 Orsay, 
Cedex, France \\
(2) Observatorio Pierre Auger, Av. San Mart\'{i}n Norte 304, (5613)
Malarg\"{u}e, Argentina} \
}

\section*{Abstract}

The detectors of the surface array of the Pierre Auger Observatory are water 
Cherenkov tanks. The signals from each tank are read out using three 
photomultipliers. The energy 
of the primary particle is inferred from signal densities and 
requires good linearity of the PMTs and a large dynamic range.
The absolute time of arrival of the shower front at each tank is obtained from 
the Global Positioning System (GPS) with a resolution of about 10 ns,
ensuring an accurate primary angular  reconstruction.
 Additionally, it is intended to use the rise time and shape of
the signals to constrain the nature of the primary particle: this sets
further requirements on the signal processing. In this paper, the main
features of the signal processing associated with the surface detector
will be presented and its performance will be discussed in the context
of the extraction of shower parameters.

\section{Introduction}

The Cherenkov light from the water tanks of the Pierre Auger Observatory is read 
out by three large photomultipliers (PMTs). The 9" XP1805 Photonis PMTs have been 
chosen for the production array. PMTs are equipped with a resistive base having 
two outputs: anode and amplified last dynode. The charge ratio between the two 
outputs is 32. This allows a large dynamic range, 
extending from few photoelectrons (pe) to about $10^{5}$ pe. The high voltage (HV)
is provided locally by a custom made ETL low power consumption (max 0.5 W) module.
The nominal operating gain of the PMTs is 2x$10^{5}$ and can be extended to 
$10^{6}$. The base, together with the HV module, is protected against humidity 
by silicone potting. 

The signals from anode and dynode are filtered with a 5 pole 
anti-aliasing Bessel filter with a cut-off at 20 MHz, and digitized at 40 MHz 
using 10 bit Flash Analog-Digital converters (FADC). The total dynamic range is 
15 bits with an overlap of 5 bits between the anode and dynode outputs.
Each detector station has 
a digital trigger circuitry implemented with Altera ACEX100 Programmable Logical
Devices (PLD), an IBM 403 PowerPC micro-controller and memory for data storage. 
The time information is obtained from the Global Positioning Satellite (GPS) 
system. The station electronics, except for the FE board, is implemented on a 
single board, called the Unified Board (UB). The FE board is plugged in the UB with 
a connector. The ensemble of electronics is packed in an aluminum enclosure. 
The electronics is mounted on top of the hatch cover of one of the PMTs and 
protected against rain and dust by an aluminum dome. In addition, each station 
has a Light Emitting Diode (LED) monitoring systems with two LEDs  (L-7113NBC 
with dominant wavelength of 445 nm) 
placed on a window on top of the liner, in the middle of the tank. The ensemble of 
electronics is powered by solar energy limited to 10 W by detector station. 
A more detailed description of the Surface Detector electronics can be found 
in [1,2].

The Surface Detector "Engineering Array" (EA) consisting of 32 water 
tanks has been in operation since June 2001 and the next phase, pre-production, 
consisting of 100 tanks with production PMTs and electronics is currently being 
deployed. The performance of the surface detector electronics has been
demonstrated by simulations, by analysis of the EA data and by
additional studies performed in various test tanks. In the following,
this performance will be discussed in the context of extraction of
shower parameters.                                                                                                       

\section{ Dynamic range and linearity }

The energy of the primary particle is inferred using the signal densities from 
a range of distances, including those from close (below 1000 m) to the shower core.
  Based on simulations, the maximum peak current at the photocathode that should 
be considered is about 250 nA. To have a linear PMT response up to this value,
 a low operating gain, 2x$10^{5}$, is required yielding a maximum current of about
 50 mA for the anode. The calibration is based on the measurement of the 
background muon spectrum. The maximum of the muon charge spectrum is located at 
the lower part of the dynamic range (dynode output). To infer the absolute 
calibration for the whole dynamic range a very good linearity  
is required for the PMTs and the electronics. 

The performance of the PMTs and of their bases has been studied by using the 
Orsay test tank. This tank is similar to the Auger tanks and has
identical LED  system [3].  The linearity of a PMT can be estimated  by varying 
the light intensity of the two LEDs (A and B) and plotting the
ratio of the  measured charges: (f(A+B)-f(A)-f(B))/f(A+B).
The PMTs linearity is better than about 
2\% up to 60 mA anode current. The specifications for the PMTs were set to better 
than 5\% linearity up to 50 mA and are confirmed in systematic tests performed 
for all PMTs by the Auger collaboration. The LED monitoring system implemented in 
each tank can be used to recheck the linearity of the PMTs and electronics.

\section{Event timing}

A common time base is established for different detector stations by using the 
GPS system. Each tank is equipped with a commercial GPS receiver (Motorola 
OnCore UT) providing a one pulse per second output and software corrections. 
This signal is used to synchronize a 100 MHz clock which is recorded at the 
time of each trigger. The intrinsic resolution of the system is about 8 ns 
requiring a good precision for the station location. An accuracy of one meter 
is obtained for the tank position by using the differential method. The time 
resolution has been studied in the EA by comparing the time signals from two 
tanks, Carmen and Miranda, separated by 11 m. A Gaussian distribution with a 17 ns 
width was obtained. This resolution combines the dispersion due to 
arrival angles  (15 ns) and the timing resolution of each tank and confirms the 
expected intrinsic time dispersion of about 8 ns. 

To perform an angle reconstruction, the starting points of the FADC traces need 
to be determined. This can be done with a resolution of about 7 ns due to the 
25 ns FADC bins. Therefore, a total time resolution of 10 ns (sigma) is achieved. 
This time resolution is sufficient to obtain a good angular resolution, 
better than one degree,  for the incoming primaries.

\section{ Signal shaping }

The discrimination of the primary composition can be inferred from the ground 
observables related to the age of the shower, hadronic heavy primaries yielding 
showers with larger muon content and earlier development.  Several estimators for 
the primary identification are currently being studied: the muon counting in the 
FADC traces, the analysis of the overall shape of the traces and the curvature 
of the shower front. 
The typical rise time for a muon signal is about 15 ns and decay time
of the order of 80 ns. The electromagnetic signals far from the shower core are 
small, of the order of a few photoelectrons. It is important that the shaping of 
the signal preserves the characteristic rise times but is also sufficient not 
to lose the small electromagnetic component of the trace in the
sampling.

\begin{figure}[t]
  \begin{center}
    \includegraphics[height=13.5pc]{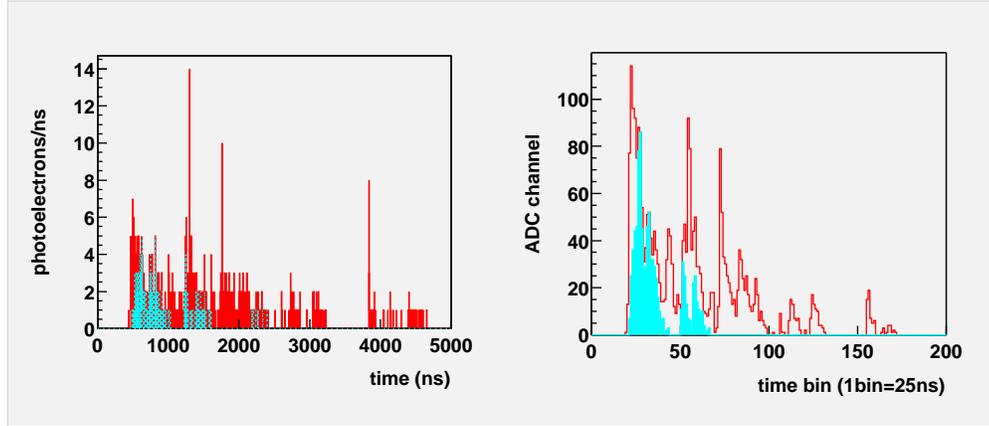}
  \end{center}
  \vspace{-0.5pc}
  \caption{ Simulated tank signal in photoelectrons per ns at the PMT photocathode (left) 
and simulated, fully processed signal (right). The muon content is
    indicated by dashed of filled area.}
\end{figure}

 Figure 1 compares a 
simulated tank signal to a simulated processed signal including  
filter and digitization effects.
Two, nearly vertical muons can be seen at around 2000 ns at the photocathode.
 After the filtering and digitization the amplitude of each muon peak
 is at around 
30 channels in the dynode output. The signals after 2000 ns are at the level 
of one or two photoelectrons and correspond to electromagnetic component only. 
These signals can be found with an amplitude of around 2-4 channels in the FADC 
trace. The integrated ratio electromagnetic over total signal is  preserved 
in the signal processing. In the case of the example of Fig.1, even in the
 late part of the signal where the electromagnetic component is small
(after 2500 ns or 100
 time bins) less than 10 \% is lost in the signal processing. Therefore, systematic errors due to electronics in the 
primary identification are estimated to be negligible. 

\section{Conclusions}

The performance of the surface detector electronics has been analyzed in the 
context of the extraction of shower parameters. The combination of two output 
signals, anode and  last, amplified dynode, covers a large dynamic range (15 bits). 
 This allows 
measurement of shower densities close to the core to extract the primary energy 
as well as to detect background muons and electromagnetic particles leaving only 
weak signals in the tank. The PMTs are linear up to anode currents larger than 
50 mA which allows 
to infer accurately the calibration from the background muons for the whole dynamic range. 
An excellent time resolution, 10 ns, is obtained allowing an accurate angle 
reconstruction. Finally, the shaping and digitizing electronics preserve the weak 
electromagnetic signals far from shower core allowing to extract primary 
composition from the trace analysis.

\vspace{1.5pc}
\re
1.\ Suomij\"{a}rvi T., Pierre Auger Collaboration,\\ 27th International Cosmic Ray
 Conference, Contributed papers vol. 1, session HE contents, 2001, 756, 
Copernicus Gesellschaft, 27. International Cosmic Ray Conference, ICRC, 
Hamburg, 07-15/08/2001 and references therein.
\re
2.\ Szadkowski Z.  \ et al. \ Contribution to this conference. 
\re
3.\ Aynutdinov V. M. \ et al.\ Contribution to this conference.

\endofpaper
\end{document}